\documentclass[aps,prl,twocolumn,longbibliography,groupedaddress,showpacs,floatfix,superscriptaddress]{revtex4-1}
\usepackage{graphicx}  
\usepackage{dcolumn}   
\usepackage[english]{babel}
\usepackage{bm}        
\usepackage{amsmath, amsthm, amssymb}
\usepackage{bbold} 
\usepackage{mathrsfs}
\usepackage{array}
\usepackage[usenames]{color}
\usepackage{color}
\usepackage{soul} 
\usepackage{ulem} 

\emergencystretch=\maxdimen
\begin{document}

\title{Phase diagram of the two-dimensional Hubbard-Holstein model: enhancement of $s$-wave pairing between charge and magnetic orders}
%competition between charge and magnetic order}
\author{Natanael C. Costa} 
\email{ndecarva@sissa.it}
\email{natanael@if.ufrj.br}
\affiliation{International School for Advanced Studies (SISSA),
Via Bonomea 265, 34136, Trieste, Italy}
\author{Kazuhiro Seki}
\affiliation{
  Computational Quantum Matter Research Team, RIKEN, Center for Emergent Matter Science (CEMS), Saitama 351-0198, Japan}
\author{Seiji Yunoki}
\affiliation{
  Computational Quantum Matter Research Team, RIKEN, Center for Emergent Matter Science (CEMS), Saitama 351-0198, Japan}
\affiliation{
  Computational Materials Science Research Team, RIKEN Center for Computational Science (R-CCS),Kobe, Hyogo 650-0047, Japan}
\affiliation{
  Computational Condensed Matter Physics Laboratory, RIKEN Cluster for Pioneering Research (CPR),Wako, Saitama 351-0198, Japan}
\author{Sandro Sorella}
\affiliation{International School for Advanced Studies (SISSA),
Via Bonomea 265, 34136, Trieste, Italy}

\begin{abstract}
We investigate the role of electron-electron and electron-phonon interactions in strongly correlated systems by performing unbiased quantum Monte Carlo simulations in the square lattice Hubbard-Holstein model at half-filling.
We study the competition and interplay between antiferromagnetism (AFM) and charge-density wave (CDW), establishing its very rich phase diagram.
In the region between AFM and CDW phases, we have found an enhancement of superconducting pairing correlations, favouring (nonlocal) $s$-wave pairs.
Our study sheds light over past inconsistencies in the literature, in particular the emergence of CDW in the pure Holstein model case.
\end{abstract}

%\date{Version 2.3 -- \today}

\maketitle

%%%%%%%%%%%%%%%%%%%%%%%%%%%%%%%%%%%%%%%%%%%%%%%%%%%%%%%%%%%%%%%%%%
%\section{Introduction}
%%%%%%%%%%%%%%%%%%%%%%%%%%%%%%%%%%%%%%%%%%%%%%%%%%%%%%%%%%%%%%%%%%
\noindent
\underbar{Introduction:}
The electron-phonon ({\it e-ph}) interaction is a central issue in condensed matter, in particular when discussing properties of conventional superconductivity (SC) and charge ordering~\cite{Giustino17}.
While Bardeen, Cooper and Schrieffer used this interaction in their seminal work to explain pairing~\cite{Bardeen57}, Peierls took it into account to provide a mechanism, based on Fermi surface nesting (FSN), that leads to charge-density wave (CDW)~\cite{peierls55}.
Recently, the debate about the role of the {\it e-ph} coupling has been intensified due to the occurrence of unconventional (non Peierls-like) CDW phases, and their competition with SC, in some classes of materials,
such as
%manganites~\cite{Calvani98,Rao00,Milward05,Cox08,Nucara08}, and
transition-metal dichalcogenides~\cite{CastroNeto01,Rossnagel11,Joe14,Ritschel15,Manzeli17,Chen16}.
%which nature is still unclear.
Even for cuprates, materials known by their strong electron-electron ({\it e-e}) interactions,
%in the $d$-orbitals of copper,
recent findings provided evidence for the occurrence of CDW in the doped region, with competing effects with SC~\cite{Lanzara01,Wise08,Ghiringhelli12,dasilvaneto14,Chen16,Achkar16},
%; e.g.,
%the critical temperatures of La$_{2-x}$Ba$_{x}$CuO$_{4}$~\cite{Ido96} and
%YBa$_{2}$Cu$_{3}$O$_{6+x}$~\cite{Kim18,Huang18,Souliou18,CyrChoiniere18}
%are enhanced when pressure is applied, as a result of the suppression of CDW correlations,
%while, conversely, the application of a magnetic field favors charge modulations~\cite{CyrChoiniere18,Wu11,Chang12}.
e.g., on doped LBCO and YBCO~\cite{Ido96,Kim18,Huang18,Souliou18,CyrChoiniere18,Wu11,Chang12}.
These results have suggested that the phase diagram of high-$T_{c}$ superconductors~\cite{Keimer15}
is far more complex than previously supposed, and have raised issues about the relevance of the {\it e-ph} coupling for correlated materials, rather than just {\it e-e} interactions.

\begin{figure}[t]
\centering
\includegraphics[scale=0.28]{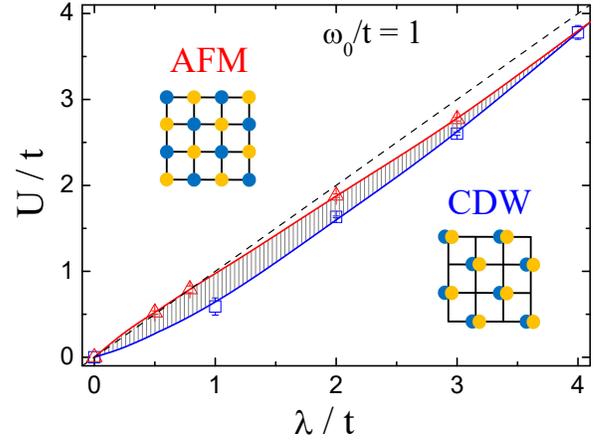}
\caption{Ground state phase diagram of the Hubbard-Holstein model at the half-filling of the square lattice, and adiabaticity ratio $\omega_{0}/t=1$. The hatched region exhibits a correlated metallic (or superconducting) behavior. The dashed line defines $U=\lambda$, while the solid lines are guide to the eyes. Here, and in all subsequent figures, when not shown, error bars are smaller than symbol size.}
\label{fig:phase_diagram} 
\end{figure}

From a theoretical point of view,
a simplified Hamiltonian that captures the interplay between antiferromagnetism (AFM), CDW, and SC is the single-band Hubbard-Holstein model (HHM)~\cite{Berger95}.
It exhibits Coulomb repulsion between electrons, leading to spin fluctuations; and also electron-ion interactions, which enhance charge/pairing correlations.
The emergence of long-range order depends on the competition between these tendencies.
This model was vastly studied in one-dimensional systems, with well-known phase diagrams presenting spin-density wave, bond-order-wave, CDW, and also metallic or phase separation behavior   ~\cite{Ning06,Matsueda06,Hohenadler13,Chakraborty14,Nocera14,Bakrim15,Lavanya17,Li18b,Hebert19,Xiao19}.
A remarkable feature in 1D systems is the occurrence of a quantum phase transition from a metallic Luther-Emery liquid phase to a CDW insulator at a finite critical {\it e-ph} coupling,
in the limit case of the pure Holstein model ($U=0$), despite of the FSN~\cite{Jeckelmann99,Hohenadler2018}.
By contrast, the properties of the HHM in two-dimensional systems are not entirely clear, even for simple geometries, such as the square lattice.
For instance, the existence of such a metal-CDW quantum critical point (QCP) is matter of controversies for the pure Holstein model in 2D lattices~\cite{Zhang19,Chen19,Hohenadler19,Weber18,Karakuzu17,Ohgoe17}.
%a critical \textit{e-ph} coupling was recently shown for the honeycomb lattice~\cite{Zhang19,Chen19}, but is matter of controversies for the square lattice case~\cite{Hohenadler19,Weber18,Karakuzu17,Ohgoe17}.
The scenario is much less clear in presence of a repulsive Hubbard term ($U\neq 0$),
in spite of the large effort to characterize the model
~\cite{Berger95,Weber18,Karakuzu17,Ohgoe17,Alder99,Khatami08,Barone08,Bauer10,Nowadnick12,
Murakami13,Nowadnick15,Pradhan15,Wang15,Mendl17},
since no unbiased results are available for quantum AFM/CDW transitions, to the best of our knowledge.

In view of these open issues, and as a step towards a better understanding of the role of {\it e-ph} interactions in strongly interacting systems, we investigate in this Letter the competition between AFM and CDW in the square lattice HHM at half-filling, as well as its pairing response, using unbiased quantum Monte Carlo (QMC) simulations.
We determine precise critical points for the HHM at intermediate interaction strengths, presenting benchmarks for lattices with linear size up to $L=64$ (i.e., 4096 sites) in some cases.
%Our results shed light over recent contronversies, with
Our main results are summarized in the ground state phase diagram displayed in Fig.\,\ref{fig:phase_diagram}.
Here we highlight 
[i] the absence of a finite critical {\it e-ph} coupling for the pure Holstein model, i.e., the CDW phase sets in for any $\lambda>0$ (and $U=0$);
[ii] the existence of a finite AFM critical point on the line $U=\lambda$, which is strongly dependent on the phonon frequency;
and [iii] an enhancement of nonlocal $s$-wave pairing in the region between the AFM and CDW phases.
These results are also compared with other methodological approaches, such as variational QMC.

%%%%%%%%%%%%%%%%%%%%%%%%%%%%%%%%%%%%%%%%%%%%%%%%%%%%%%%%%%%%%%%%%%
%\section{Model and Method} \label{Model and Method}
%%%%%%%%%%%%%%%%%%%%%%%%%%%%%%%%%%%%%%%%%%%%%%%%%%%%%%%%%%%%%%%%%%

\noindent
\underbar{Methodology:}
The Hubbard-Holstein Hamiltonian reads
\begin{align} \label{eq:HHM_hamil}
\nonumber \mathcal{H} = &
-t \sum_{\langle \mathbf{i}, \mathbf{j} \rangle, \sigma} 
\big(d^{\dagger}_{\mathbf{i} \sigma} d^{\phantom{\dagger}}_{\mathbf{j} \sigma} + {\rm h.c.} \big)
- \mu \sum_{\mathbf{i}, \sigma} n^{\phantom{\dagger}}_{\mathbf{i}, \sigma}
+ U \sum_{\mathbf{i}} n_{\mathbf{i}\uparrow} n_{\mathbf{i}\downarrow}
\\
&
+ \sum_{ \mathbf{i} }
\bigg( \frac{\hat{P}^{2}_{\mathbf{i}}}{2 M} + \frac{M \omega^{2}_{0} }{2} \hat{X}^{2}_{\mathbf{i}}\bigg)
- g \sum_{\mathbf{i}, \sigma} n_{\mathbf{i}\sigma} \hat{X}_{\mathbf{i}}
,
\end{align}
where $d^{\dagger}_{\mathbf{i} \sigma}$ ($d^{\phantom{\dagger}}_{\mathbf{i} \sigma}$)
is a creation (annihilation) operator of electrons with spin $\sigma\,(=\uparrow,\downarrow)$ at a given site $\mathbf{i}$ 
on a two-dimensional square lattice under periodic boundary conditions, with $\langle \mathbf{i}, \mathbf{j} \rangle$ denoting 
nearest-neighbors, and 
$n^{\phantom{\dagger}}_{\mathbf{i}\sigma}\equiv d^{\dagger}_{\mathbf{i} \sigma} d^{\phantom{\dagger}}_{\mathbf{i} \sigma}$
being number operators.
The first two terms on the right hand side of Eq.\,\eqref{eq:HHM_hamil} correspond to the kinetic energy of electrons, and their chemical potential ($\mu$) term, respectively, while the on-site Coulomb repulsion between electrons is included by the third term.
%The vibration of ions is
The ions' phonon modes are
described in the fourth term, in which $\hat{P}_{\mathbf{i}}$ and $\hat{X}_{\mathbf{i}}$ are momentum and position operators, respectively, of local quantum harmonic oscillators with frequency $\omega_{0}$.
The last term corresponds to local electron-ion interactions, with strength $g$.
Hereafter, we define the mass of the ions ($M$) and the lattice constant as 
unity.

It is also worth to introduce additional parameters, due to the effects of the phonon fields to the electronic interactions.
From a second order perturbation theory on the {\it e-ph} term~\cite{Berger95}, one obtains an effective dynamic {\it e-e} interaction,
$U_{\rm eff} (\omega) =  U - \frac{g^{2} / \omega^{2}_{0}}{1 - (\omega /\omega_{0})^2 }$, with $g^{2} / \omega^{2}_{0} \equiv \lambda$ being the energy scale for polaron formation.
The appearance of such a retarded attractive interaction, depending on the phonon frequency, leads us to define $\omega_{0}/t$ as the adiabaticity ratio, and $\lambda /t$ as the strength of the {\it e-ph} interaction.
To facilitate the following discussion, we also define $U_{\rm eff} \equiv U - \lambda$, % as the effective interaction.
which gives us information about the local effective $e$-$e$ interaction, and is also an important parameter to our methodological approaches. 
Furthermore,
%except when explicitly stated, we choose $\omega_{0}/t =1$, while varying $U/t$ and $\lambda/t$,
we set the electron density at half-filling, i.e., $\langle n_{\mathbf{i}\sigma} \rangle = 1/2$.

We investigate the properties of Eq.\,\eqref{eq:HHM_hamil} by performing two different unbiased auxiliary-field QMC approaches: the projective ground state auxiliary-field (AFQMC)~\cite{Sorella89,Blankenbecler81}, and the finite temperature determinant quantum Monte Carlo (DQMC) methods~\cite{Blankenbecler81,Hirsch83,Hirsch85,Scalettar89}.
Briefly, the DQMC (AFQMC) approach is based on the decoupling of the non-commuting terms of the Hamiltonian in the partition function
(projection operator)
by Trotter-Suzuki decomposition, which discretizes the imaginary-time coordinate $\tau$ in small
intervals $\Delta\tau$, with the inverse of temperature $T$ (projection time) being $\beta=M\Delta\tau$.
%The interacting terms are transformed in one-particle operators by means of a discrete Hubbard-Stratonovich transformation, 
%with the cost of including bosonic auxiliary-fields $\mathcal{S}(\tau,\mathbf{i})$, in real space and imaginary-time coordinates, 
%coupled to fermionic degrees of freedom. Monte Carlo methodologies are used for sampling $\mathcal{S}(\tau,\mathbf{i})$; 
Throughout this Letter, we choose $\Delta\tau t=0.1$, with $\beta$ in unit of $t$.
Detailed introduction for these methodologies can be found, e.g., in Refs.\,\onlinecite{Santos03,gubernatis16,becca17}.

Following the procedures described in Ref.\,\onlinecite{Karakuzu18}, we implemented a sign-free AFQMC approach to the half-filling of the HHM, allowing us to analyze large lattice sizes, but, conversely, being restricted to the $U_{\rm eff} \geq 0 $ region.
The properties of the $U_{\rm eff} < 0$ region, forbidden to our AFQMC method, are investigated by DQMC simulations.
We recall that the DQMC method may exhibit sign problem for the HHM, depending on the strength 
of parameters.
However, the average sign is less affected when $U < \lambda$, in particular to intermediate interaction strengths, allowing us to obtain the physical quantities of interest, in some cases up to $L=14$ and $\beta = 28$.
In fact, the DQMC average sign is strongly suppressed just around $U \approx \lambda$, where our sign-free AFQMC approach works.
Thus, our AFQMC and DQMC simulations are used complementarily.

The charge and magnetic responses are quantified by their respective structure factors, i.e.,
$S_{\rm cdw}(\mathbf{q}) = \frac{1}{N} \sum_{\mathbf{i}, \mathbf{j}} e^{-{\rm i}\mathbf{q}\cdot(\mathbf{i} - \mathbf{j})} \langle n_{\mathbf{i}} n_{\mathbf{j}} \rangle $, and 
$S_{\rm afm}(\mathbf{q}) = \frac{1}{N} \sum_{\mathbf{i}, \mathbf{j}} e^{-{\rm i}\mathbf{q}\cdot(\mathbf{i} - \mathbf{j})} \langle S^{z}_{\mathbf{i}} S^{z}_{\mathbf{j}} \rangle $, with
$n_{\mathbf{i}}=n_{\mathbf{i} \uparrow} + n_{\mathbf{i} \downarrow}$,
$S^{z}_{\mathbf{i}}=n_{\mathbf{i} \uparrow} - n_{\mathbf{i} \downarrow}$, and
$N=L \times L$ being the number of sites.
This allows us to probe their critical behavior by means of the correlation ratio
\begin{align}\label{eq:Rc}
R_{\nu}(L) = 1 - \frac{S_{\nu}(\mathbf{q}+\delta\mathbf{q})}{S_{\nu}(\mathbf{q})},
\end{align}
with $|\delta\mathbf{q}|= 2\pi/L$, $\mathbf{q}=(\pi,\pi)$,
and $\nu \equiv$ cdw or afm.
According to well established finite size scaling analysis, the critical region is determined by the  
%the scaling of
crossing points of $R_{\nu}(L)$ for different lattice sizes (see, e.g., Refs.\,\onlinecite{Kaul15,Gazit18,Sato18,Liu18,Darmawan18}).
Finally, the pairing properties are investigated by the finite temperature superconducting pair susceptibility
$
\chi^{\alpha}_{\rm sc}(\beta) = \frac{1}{N} 
\int^{\beta}_{0} \mathrm{d}\tau \,
\langle \Delta_{\alpha}(\tau) \Delta^{\dagger}_{\alpha}(0) + \mathrm{H.c.} \rangle,
$
with
$ \Delta_{\alpha}(\tau) = \frac{1}{2} \sum_{\mathbf{i}, \mathbf{a}} f_{\alpha}(\mathbf{a})
c^{\phantom{\dagger}}_{\mathbf{i}\downarrow}(\tau) 
c^{\phantom{\dagger}}_{\mathbf{i}+\mathbf{a} \uparrow}(\tau)$, 
$c_{\mathbf{i}\sigma}(\tau)=e^{\tau\cal{H}}c_{\mathbf{i}\sigma}e^{-\tau\cal{H}}$
and $f_{\alpha}(\mathbf{a})$ being the 
pairing form factor for a given symmetry.
Here, we consider on-site, nearest-neighbors (NN), and next-nearest-neighbors (NNN) spin-singlet pairing operators for the $s$-wave symmetry, which are denoted by $\alpha\equiv s$, $s^*$, and $s^{**}$, respectively~\footnote{$s^{**}$ is sometimes called as $s_{xy}$ in literature.};
and also consider the $d_{x^2 - y^2}$-wave symmetry, $\alpha\equiv d$ (see, e.g., Ref.\,\onlinecite{White89b}).

%%%%%%%%%%%%%%%%%%%%%%%%%%%%%%%%%%%%%%%%%%%%%%%%%%%%%%%%%%%%%%%%%%
%\section{Results} \label{Results}
%%%%%%%%%%%%%%%%%%%%%%%%%%%%%%%%%%%%%%%%%%%%%%%%%%%%%%%%%%%%%%%%%%
\begin{figure}[t]
\centering
\includegraphics[scale=0.25]{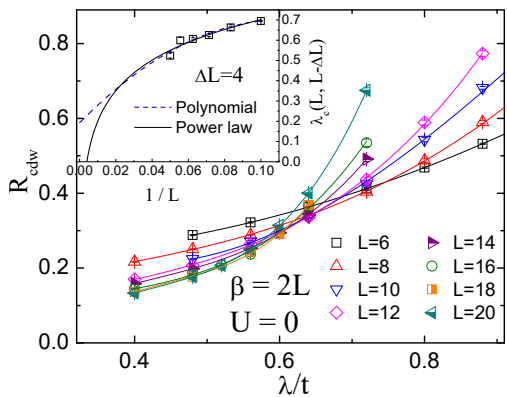}
\caption{DQMC results for the CDW correlation ratio, Eq.\eqref{eq:Rc}, as function of $\lambda/t$, for different lattice sizes, and by fixing $\beta=2L$, $\omega_{0}/t=1$, and $U=0$. The solid lines are just guide to the eyes.
Inset: $\lambda_{c}(L,L-\Delta L)$ as a function of $1/L$; the solid (dashed) curve corresponds to its power law (polynomial) scaling.
  }
\label{fig:U00lambda} 
\end{figure}

\noindent
\underbar{Results:}
We first discuss the limit case of $U=0$ in Eq.\,\eqref{eq:HHM_hamil}, i.e., the pure Holstein model.
%As mentioned earlier, the Peierls' mechanism for CDW formation is based on the occurence of FSN, since a diverging bare electronic susceptibility at the nesting wavevector eventually leads to an insulating charge ordered ground state, with $\mathbf{q}_{\rm CDW}=2\mathbf{k}_{\rm F}$, for any $\lambda>0$ -- by arguments of Random Phase Approximation~\cite{Gruner94}.
While the Peierls' argument~\cite{Gruner94}
%, based on the occurence of FSN,
suggests an insulating charge ordered ground state
%with $\mathbf{q}_{\rm cdw}=2\mathbf{k}_{\rm F}=(\pi,\pi)$,
for any $\lambda>0$, one-dimensional systems exhibit a metal-CDW transition for a finite critical $\lambda_{c}$, despite the perfect FSN.
In two dimensions,
%the square lattice also exhibits FSN at half-filling, but
the occurrence of a finite critical $\lambda_{c}$ in the square lattice is controversial: while variational QMC approaches provide evidence for $\lambda_{c}/t \approx 0.8$~\cite{Karakuzu17,Ohgoe17}, unbiased QMC results suggest the nonexistence of a finite critical point~\cite{Hohenadler19,Weber18}.
Here, we address this controversy by analyzing the critical behavior given by the CDW correlation ratio, Eq.\,\eqref{eq:Rc}, by means of DQMC simulations
\footnote{For $U=0$ the DQMC method does not suffer with the sign problem, allowing us to investigate larger lattice sizes and lower temperatures.}.

The quantum critical region may be estimated by the crossing of $R_{\rm cdw}(L)$ as function of $\lambda /t$, as displayed in Fig.\,\ref{fig:U00lambda} for several lattice sizes.
Here we
investigate the ground state behavior by
assuming $\beta \sim L^{z}$, with $z=1$ being the dynamic critical exponent.
Notice that, the correlation ratio increases monotonically as a function of the lattice size starting from $\lambda/t = 0.8$, clearly supporting long-range CDW ordering at this interaction strength and above.
Furthermore, as showed in the inset of Fig.\,\ref{fig:U00lambda},
the $\lambda_{c}(L,L-\Delta L)$, i.e.\,~the values of $\lambda/t$ for the crossing points between $R_{\rm cdw}(L)$ and $R_{\rm cdw}(L-\Delta L)$,
are reduced when $L$ increases, suggesting that, whether a metal-insulator transition exists, it should occur at smaller coupling strengths.
A thorough determination of the existence of a critical point is given by a finite size scaling analysis of
%$R_{\rm cdw}(L)$.
$\lambda_{c}(L,L-\Delta L)$.
Following the procedure adopted in Ref.\,\onlinecite{Weber18}, which hereafter is used to determine the CDW transitions, we perform a power law scaling [$f(L)= a + b L^{c}$] of the crossing points, as displayed in the inset of Fig.\,\ref{fig:U00lambda}.
Within this scaling, $\lambda_{c}$ ($U=0$) is consistent with a vanishing or very small value, even when we adopt the less accurate polynomial fit, indicating that a finite critical \textit{e-ph} coupling is not plausible for the square lattice Holstein model.
%, in line with Peierls' argument.
%\footnote{Notice that a polynomial fit (blue dashed curve) provides a critical $\lambda_{c}/t \approx 0.2$ (or $\lambda_{D} \approx 0.025$), much lower than the estimation given by variational methods.}.
The difference between the square lattice and one-dimensional systems may stem on the larger electronic susceptibility of the former, which diverges with the square logarithm of temperature.

\begin{figure}[t]
\centering
\includegraphics[scale=0.25]{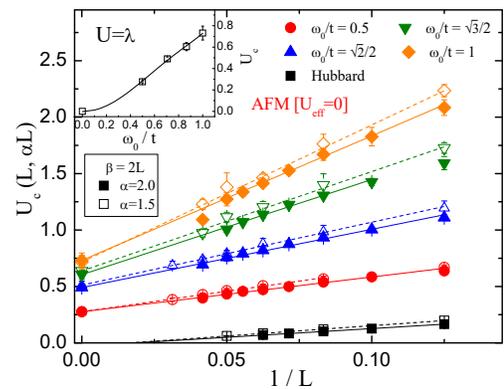}
\caption{AFQMC results for the crossing points of $R_{\rm afm}(L)$ and $R_{\rm afm}(\alpha L)$, for different phonon frequencies, and by fixing $\beta=2L$ and $U_{\rm eff}=0$. The lines are linear scalings. Inset: Critical AFM points for $U_{\rm eff}=0$ as function of the phonon frequency $\omega_{0}/t$.
}
\label{fig:Ueff} 
\end{figure}

We now discuss the behavior of the HHM for $U \neq 0$, investigating initially the particular case of $U=\lambda$, by means of AFQMC simulations.
We recall that,
the HHM in the antiadiabatic limit ($\omega_{0}\to\infty$) should exhibit a metallic behavior along the line $U=\lambda$.
%, due to an instantaneous effective interaction.
However, the occurrence of a retarded interaction in the adiabatic limit
%change the electronic response, and
leads to a more complex ground state.
Our QMC results for $\omega_{0}/t \leq 1$ (not shown) exhibit an enhancement of the spin-spin correlations as a function of $U/t$, on the line $U=\lambda$, while the charge-charge response remains weak for any interaction strength.
This result suggests an AFM ground state, with long-range order being probed by its correlation ratio.
Similarly to CDW analysis, Fig.\,\ref{fig:Ueff} displays the
$U_{c}(L,\alpha L)$, i.e. the values of $U/t$ for
the crossing points of $R_{\rm afm}(L)$ and $R_{\rm afm}(\alpha L)$,
for different phonon frequencies, and fixing $\beta=2L$.
Due to the large lattice sizes achieved in our AFQMC simulations, here we adopt a linear finite size scaling for the AFM transitions.
%; see, e.g., the solid and dashed lines in Fig.\,\ref{fig:Ueff}.
As expected, the pure Hubbard model (black square symbols in Fig.~\ref{fig:Ueff}) is AFM for any $U>0$, i.e., $U_{c}=0$.
However, in presence of {\it e-ph} coupling (along the line $U=\lambda$), a quantum phase transition occurs, changing from a correlated metallic-like ground state to an ordered AFM one, for a given coupling strength.
For instance, for $\omega_{0}/t = 1$, one finds $U_{c}/t=\lambda_{c}/t=0.73(6)$, that is, the ground state is AFM for any $U=\lambda>0.73t$.
The position of this QCP strongly depends on the choice of $\omega_{0}/t$, as showed in the inset of Fig.\,\ref{fig:Ueff}.
Such increasing behavior is consistent with the expectation of an emergent metallic behavior in the antiadiabatic limit.
The properties of this correlated metallic-like state are discussed later.

\begin{figure}[t]
\centering
\includegraphics[scale=0.30]{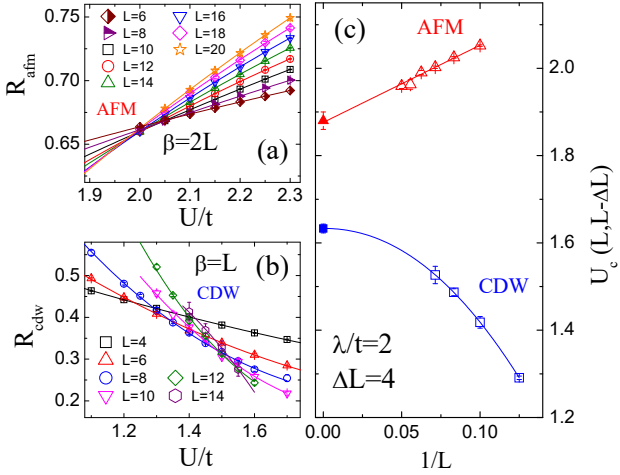}
\caption{(a) AFQMC results for the AFM correlation ratio for $\beta=2L$, and (b) DQMC results for the CDW correlation ratio for $\beta=L$, as function of the $U/t$, for different lattice sizes, and by fixing $\lambda/t=2$ and $\omega_{0}/t=1$.
The solid lines are just guide to the eyes.
(c) The crossing points of $R_{\rm afm}(L)$ (red triangles), and $R_{\rm cdw}(L)$ (blue squares), and their respective scaling curves.
}
\label{fig:lambda20} 
\end{figure}

%\begin{figure}[b]
%\centering
%\includegraphics[scale=0.24]{Tcdw_lbD40_v1.pdf}
%\caption{CDW critical temperature as function of $U/t$, for fixed $\omega_{0}/t=1$, and $\lambda/t=4$ ($\lambda_{D}=0.50$). The solid line is just guide to the eyes.}
%\label{fig:Tc_cdw} 
%\end{figure}

We proceed by investigating the quantum critical behavior for the general case, $U \neq \lambda$.
To this end, one may analyze the correlation ratio as function of $U/t$, for fixed $\lambda /t$ and $\omega_{0}/t$.
For instance, Fig.\,\ref{fig:lambda20}\,(a) displays AFQMC results for $R_{\rm afm}(L)$ as function of $U/t$, for $\lambda /t=2$ and $\omega_{0}/t=1$.
The crossing points $U_{c}(L,L-\Delta L)$ between $R_{\rm afm}(L)$ and $R_{\rm afm}(L-\Delta L)$, as well as their finite size scaling, are displayed in Fig.\,\ref{fig:lambda20}\,(c), leading to an AFM quantum phase transition at $U_{c}^{\rm AFM}/t = 1.88(2)$.
Similarly, the CDW QCP may be obtained by DQMC simulations of $R_{\rm cdw}(L)$, as presented in Fig.\,\ref{fig:lambda20}\,(b), with crossing points and finite size scaling shown in Fig.\,\ref{fig:lambda20}\,(c), leading to $U_{c}^{\rm CDW}/t = 1.63(1)$.

When the above analysis is repeated for other values of $\lambda /t$, we obtain the phase diagram presented in Fig.\,\ref{fig:phase_diagram}.
It is worth mentioning that, for the range of parameters analyzed, we obtain continuous transitions for both AFM and CDW phases, without coexistence, and with a metallic-like (or SC) region between them.
%For instance, by analyzing the CDW critical temperatures for $\lambda/t=4$, as well as the behavior of $S_{\rm cdw}(\mathbf{q})$, our results are consistent with a continuous phase transition; see, e.g., the Supplemental Material.
%see, e.g., the Supplemental Material for results with $\lambda/t=4$.
First order transitions may occur for stronger coupling, as suggested in Refs.\,\onlinecite{Karakuzu17,Ohgoe17}.

\begin{figure}[t]
\centering
\includegraphics[scale=0.30]{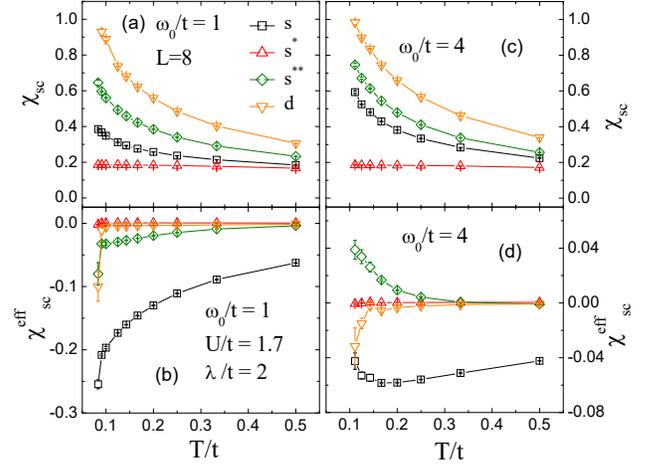}
\caption{DQMC results for the pair susceptibility as function of temperature $T$, for fixed $\lambda/t=2$, $U/t=1.7$, $L=8$, and (a) $\omega_{0}/t=1$ and (c) $\omega_{0}/t=4$, and their respective effective susceptibilities for (b) $\omega_{0}/t=1$ and (d) $\omega_{0}/t=4$.
}
\label{fig:SC} 
\end{figure}

Finally, it is instructive to discuss the properties of such a region between AFM and CDW phases, from which it is expected the emergence of SC.
%In absence of long-range spin and charge orders, one expects the emergence of SC.
% naively $s$-wave-like, due to phonon-mediated interactions.
Although establishing long-range SC order is challenging, its properties may be investigated from the tendency of the pairing susceptibility as a function of temperature.
For instance, by using the DQMC method, and
fixing $\lambda/t=2$, and $U/t=1.7$, we observe an enhancement of $\chi_{\alpha \rm (sc)}$ at low temperature, with the dominant symmetry being the $d$-wave~\cite{Mendl17}, as presented in Fig.\,\ref{fig:SC}\,(a).
However, a more appropriate quantity for SC is given by
extracting the particle-particle contribution of $\chi_{\alpha \rm (sc)}$, which defines the effective pair (vertex) susceptibility, i.e., $\chi^{\rm eff}_{\alpha {\rm (sc)}} = \chi_{\alpha {\rm (sc)}} - \bar{\chi}_{\alpha {\rm (sc)}}$, with $\bar{\chi}_{\alpha {\rm (sc)}}$ being the
%susceptibility of single-particle (noninteracting) propagators
noninteracting susceptibility~\cite{White89b}.
A positive (negative) response of $\chi^{\rm eff}_{\alpha {\rm (sc)}}$ corresponds to an enhancement (weakening) of pair correlations for the $\alpha$th symmetry.
Figure \ref{fig:SC}\,(b) exhibits $\chi^{\rm eff}_{\alpha {\rm (sc)}}$ for the data of panel (a), showing negative tendency towards all of the examined channels.
In particular, the on-site $s$-wave has the largest negative response,
%suggesting that 
which shows the harmfulness of the Hubbard-like term for local pairs formation.
%Since $T_{sc} \sim \omega_{0} e^{-\omega^{2}_{0}/\lambda N(0)}$
Since $T_{sc} \sim \omega_{0}$ from the BCS theory~\citep{Bardeen57}, further insights about the nature of this region may be given by increasing $\omega_{0}$, while keeping $U_{\rm eff}$ fixed, as displayed in Figs.\,\ref{fig:SC}(c) and \ref{fig:SC}(d), for $\omega_{0}/t=4$, $\lambda/t=2$, and $U/t=1.7$.
For these parameters, despite the increasing dominant behavior of $\chi_{\alpha \rm (sc)}$ for the $d$-wave, only the NNN $s$-wave exhibits a positive effective susceptibility.

It is important to mention that, despite the dominant character of the $d$-wave in $\chi_{\alpha \rm (sc)}$, for both adiabatic and antiadiabatic limits, its negative tendencies for $\chi^{\rm eff}_{\alpha {\rm (sc)}}$ suggest that such a channel may not lead to long-range SC order in the ground state of the half-filled HHM.
Conversely, whether SC emerges, the results of Fig.\,\ref{fig:SC} provide evidence for (nonlocal) $s$-wave.
Indeed, since this metallic-like region is mostly in the negative $U_{\rm eff}$ side of the phase diagram in Fig.\,\ref{fig:phase_diagram}, the $s$-wave symmetry seems more plausible than $d$-wave.
Interestingly, such an enhancement of nonlocal $s$-wave response suggests that short-range charge/spin correlations may suppress the formation of local ($s$-wave) and NN ($s^{*}$-wave) Cooper pairs, making the NNN ones ($s^{**}$-wave) the main channel for pairing.

%%%%%%%%%%%%%%%%%%%%%%%%%%%%%%%%%%%%%%%%%%%%%%%%%%%%%%%%%%%%%%%%%%
%\section{Conclusions} \label{Conclusions}
%%%%%%%%%%%%%%%%%%%%%%%%%%%%%%%%%%%%%%%%%%%%%%%%%%%%%%%%%%%%%%%%%%
\noindent
\underbar{Conclusions:}
We have presented results for the HHM in the square lattice, using unbiased AFQMC and DQMC methods 
complementarily, which provide a broader picture about the physical responses of this model.
In particular, we have shown that, different from one-dimensional systems, the emergence of the CDW phase in the square lattice occurs for any $\lambda > 0$, for $U=0$.
However, these CDW correlations are strongly affected by a Coulomb interaction ($U \neq 0$), with the occurrence of AFM even at $U_{\rm eff} < 0$.
We also observed the existence of a correlated metallic-like region between CDW and AFM phases, with an enhancement of nonlocal $s$-wave pairing, rather than $d$-wave.
Despite the difficulty to establish long-range order for SC, one may expect that $s$-wave SC sets in at zero temperature.
%Despite the difficulty to establish long-range order for SC, it is worthy mentioning that the on-site $s$-wave pairing, the most analyzed pairing for SC in the HHM literature, is indeed the less likely one, having their pair correlation functions weakened.
%However, our results are consistent that nonlocal $s$-wave pairing is enhanced, instead of $d$-wave.
These findings constitute a significant step towards a better understanding of this model, by providing precise QCPs, and shedding lights over past theoretical inconsistencies.
Furthermore, these results emphasize the role of the \textit{e-ph} interaction in strongly correlated systems,
which is crucial to charge order and pairing,
and may be relevant for the physics of cuprates and transition-metal dichalcogenides.

%%%%%%%%%%%%%%%%%%%%%%%%%%%%%%%%%%%%%%%%%%%%%%%%%%%%%%%%%%%%%%%%%%
%\section*{ACKNOWLEDGMENTS}
%%%%%%%%%%%%%%%%%%%%%%%%%%%%%%%%%%%%%%%%%%%%%%%%%%%%%%%%%%%%%%%%%%
\begin{acknowledgments}
\underbar{Acknowledgements:}
We are grateful to Yuichi Otsuka for valuable discussions.
Computational resources were provided by HOKUSAI supercomputer at RIKEN (Project ID:~G19010),
and CINECA supercomputer (PRACE-2019204934).
S.S.~and N.C.C.~acknowledge PRACE for awarding them access to Marconi at CINECA, Italy.

\end{acknowledgments}

%%%%%%%%%%%%%%%%%%%%%%%%%%%%%%%%%%%%%%%%%%%%%%%%%%%%%%%%%%%%%%%%%%%%%%%%
%%%
%%%%%     BIBLIOGRAPHY
%%%%%%%%%%%%%%%%%%%%%%%%%%%%%%%%%%%%%%%%%%%%%%%%%%%%%%%%%%%%%%%%%%%%%%%%%%%

\bibliography{bibCostaHH}

\end{document}